\documentclass[12pt]{article}
\usepackage{wider}
\begin{document}

\title{\bf A Quantum Search Algorithm for a Specified Number of Targets\thanks{
This work was sponsored by the Department of the Air Force under Contract 
F19628-00-C-0002.  Opinions, interpretations, conclusions, and 
recommendations are those of the author and are not necessarily endorsed 
by the United States Air Force.
}
}
\author{
Mark A. Rubin\\
\mbox{}\\   
Lincoln Laboratory\\ 
Massachusetts Institute of Technology\\  
244 Wood Street\\                         
Lexington, Massachusetts 02420-9185\\      
{\tt rubin@ll.mit.edu}\\ 
}
\date{\mbox{}}
\maketitle

\begin{abstract}

The quantum search algorithm of Chen and Diao, which
finds with certainty  a single target item in an unsorted database, is modified so as to
be capable of searching for an arbitrary specified number of target items.
If the  number of targets, $\nu_0$\/, is a power of four, the new algorithm will 
with certainty 
find one of the targets  in a database of ${\cal N}$\/ items 
using  
$(1/2)\left(3(N/\nu_0)^{\log_43}-1\right)
\approx
(1/2)\left(3(N/\nu_0)^{0.7925}-1\rule{0cm}{.35cm}\right)$\/
oracle
calls, where $N$\/ is the smallest  power of four greater than or equal to 
${\cal N}$\/.   
If $\nu_0$\/ is not a power of four, the algorithm will, 
with a probability of at least one-half,
find
one of the targets  
using no more than 
$(1/2)\left(9(N/\nu)^{\log_43}-1\right)$\/
calls, where $\nu$\/ is the smallest  power of four greater than or equal to 
$\nu_0$\/.

\end{abstract}

\section{Introduction}

Recently Chen and Diao \cite{ChenDiao00}  presented a quantum algorithm for
searching an unsorted database   capable of finding, with certainty,
a single target item in an ${\cal N}$\/-item database    after  $2 \lceil 
\log_4{\cal N} \rceil$\/ iterations of  certain unitary operations. 
($\lceil x \rceil $\/ denotes the smallest integer greater than or equal to $x$\/.)
Grassl \cite{Grassl01} and Tu and Long \cite{TuLong01} have given a recursive implementation of these unitary operations, and have
pointed out that, with this implementation, the number of oracle calls
required for the $j^{th}$\/ iteration  increases
exponentially with $j$\/.

In this paper I present a modification of the algorithm of \cite{ChenDiao00}
for searching an unsorted database of ${\cal N}$\/ items for $\nu_0 \ge 1$\/
target items, provided that the number of targets $\nu_0$\/
is known in advance.  In Section 2 below I discuss the case of $\nu_0$\/
equal to a power of four; in this case  the algorithm will
find one of the target items with unit probability.  In Section 3 I discuss the
case  of $\nu_0$\/ not equal to a power of four; in this case  the algorithm will
find one of the target items with probability  of at least
one-half. The number of oracle calls required using the recursive  
implementation is given in Section 4. The notation and terminology follow, in general,  those of \cite{ChenDiao00} and \cite{Grassl01}.

\section{Number of Targets a Power of Four}

Denote the ${\cal N}$\/ items in the database ${\cal D}$\/ by $w_i, i=1,\ldots, {\cal N}$\/.
Of these items, a total of $\nu_0$\/ are members of the subset $T$\/ of target items.
An oracle function $f(w_i)$\/
indicates whether
a selected item is or is not a target:
\begin{equation}
\begin{array}{lcl}
f(w_i)& = 1, & \;  w_i \in T,\\
      & = 0, & \; \mbox{\rm otherwise.} \label{oracle}
\end{array}
\end{equation}
If ${\cal N}$\/ is not already a power of four, we embed the database ${\cal D}$\/
in a larger database $D$\/ containing additional  non-target items such that the total 
number of items in $D$\/ is the smallest power of four larger than ${\cal N}$\/:
\begin{equation}
D={\cal D} \cup \{w_{{\cal N}+1},\ldots, w_{N}\},
\end{equation}
where 
\begin{equation}
N=2^{2n},
\end{equation}
$n$\/ an integer, i.e.,
\begin{equation}
n= \lceil \log_4 {\cal N}\/  \rceil ,
\end{equation}
so
\begin{equation}
N > {\cal N} > N/4.
\end{equation}

The above enlargement of the database is as in \cite{ChenDiao00}.  Here, in addition,
we embed $D$\/ in a database $\widetilde{D}$\/ which is four times larger still:
\begin{equation}
\widetilde{D}=D \cup \{w_{N+1}, w_{N+2}, \ldots, w_{\widetilde{N}}\},
\end{equation}
where
\begin{equation}
\widetilde{N}=4N=2^{2\tilde{n}}.
\end{equation}
That is,
\begin{equation}
\tilde{n}=n+1.
\end{equation}
All of the additional items not in ${\cal D}$\/ are by definition 
non-targets, so equation (\ref{oracle})
still holds and the cardinality of $T$\/ is still $\nu_0$\/.

For the database to be searched by a quantum 
computer \cite{MerminQCnotes99}, the $\widetilde{N}$\/
items in $\widetilde{D}$\/ are set in one-to-one correspondence with the $\widetilde{N}$\/
computational-basis states $|a_1 a_2 \ldots a_{2\tilde{n}}\rangle$\/:
\begin{equation}
w_i \leftrightarrow |a_1(i) a_2(i)  \ldots a_{2\tilde{n}}(i)\rangle, \label{correspondence}
\;\;\; i=1,\ldots,  \widetilde{N}
\end{equation}
where each of the eigenvalues $a_j(i)$\/  is either 0 or 1. The $2\tilde{n}$\/-component vector of $a_j$\/'s associated with 
$w_i$\/ is termed the symbol of $w_i$\/:
\begin{equation}
S(w_i)=a_1(i) a_2(i) \ldots a_{2\tilde{n}}(i).
\end{equation}
We also define auxiliary symbol functions
\begin{equation}
\begin{array}{lcl}
S_{j}(w_i)&=& a_1(i)\ldots a_j(i),\; \; \; j=1,\ldots,2\tilde{n},\\
S_{2\tilde{n}}(w_i)&=& S(w_i). \label{auxiliarysymbol}
\end{array}
\end{equation}

It should be emphasized that the  correspondence (\ref{correspondence}) 
is {\em not}\/ chosen to make the symbol $S(w_i)$\/ a binary representation
of the item index $i$\/. On the contrary, it is essential for what follows
that none of the $N$\/ items in the set $D$\/ be represented by states such that
$S_2(w_i)=00$\/. That is, we require that
\begin{equation}
w_i \in D \Rightarrow S_2(w_i) \neq 00. \label{notin00}
\end{equation}
(We could, for example, establish the  correspondence (\ref{correspondence})
so that $w_i \in D \Rightarrow
S_2(w_i)=11$\/.) 
Condition
(\ref{notin00}) implies 
\begin{equation}
w_i \in T \Rightarrow S_2(w_i) \neq 00.
\end{equation}

Extending the technique employed in \cite{ChenDiao00} to the case of multiple
targets, we select  $\nu_0$\/
of the items with auxiliary symbols $S_2(w_i)=00$\/ to be ``ground state items.''
Specifically, the $\nu_0$\/ elements of the set $G$\/ of ground state items,
\begin{equation}
G=\{w_{G_1}, w_{G_2},\ldots,w_{G_{\nu_0}}\},
\end{equation}
are those with the symbols
\begin{equation}
\begin{array}{lcl}
S(w_{G_1})&=& 00 \ldots 000000 \\
S(w_{G_2})&=& 00 \ldots 000001 \\
S(w_{G_3})&=& 00 \ldots 000010 \\
S(w_{G_4})&=& 00 \ldots 000011 \\
 & \vdots&  \label{Gsymbols}
\end{array}
\end{equation}
The rightmost $2p$\/ entries in $S(w_{G_{\nu_0}})$\/ are all 1's and constitute
a binary
representation of $\nu_0-1$\/, where
\begin{equation}
2^{2p}=\nu_0. \label{pdef0}
\end{equation}

We can now define the auxiliary functions
\begin{equation}
\begin{array}{lcl}
f_j(w_i)&=&1 \;\; \; \mbox{\rm if} \; S_{2j}=00 \ldots 00\;  \mbox{\rm but}\; w_i \not\in G,\\
        &=&0 \;\; \;  \mbox{\rm otherwise;} \;\; \;  j=1, \ldots, \tilde{n}-p,
\label{auxiliaryfunction}
\end{array}
\end{equation}
and, in terms of these, the auxiliary oracle functions 
\begin{equation}
\begin{array}{lcl}
F_j(w_i)=f(w_i) \vee f_j(w_i). \label{auxiliaryoracle}
\end{array}
\end{equation}
(The symbol ``$ \vee $\/'' denotes logical OR.)
Note that
\begin{equation}
F_{\tilde{n}-p}(w_i)=f(w_i). \label{Fn_p}
\end{equation}

The starting state for the iteration is the equally-weighted superposition of 
computational basis states obtained from the state 
$|w_{G_1}\rangle= |00 \ldots 00\rangle$\/ by a Walsh-Hadamard transformation,
\begin{equation}
|s_0\rangle=\frac{1}{\sqrt{\widetilde{N}}}\sum_{i=1}^{\widetilde{N}}|w_i\rangle.
\label{startingstate}
\end{equation}
Starting from $|s_0\rangle$\/, a total of $n_I^0$\/ 
iterations are performed of the transformation
\begin{equation}
|s_{j+1}\rangle = -{\cal I}_{s_j}{\cal I}_j |s_j\rangle, \; j=0, \ldots, n^0_I-1,
\label{iteration}
\end{equation}
where
\begin{equation}
n_I^0 = \tilde{n}-p.  \label{nI0}
\end{equation}
The unitary operator ${\cal I}_j$\/ in (\ref{iteration}) is defined as
\begin{equation}
{\cal I}_j= \mbox{\rm I} -2\sum_{i | F_{j+1}(w_i)=1} |w_i\rangle\langle w_i|, \label{I_j1}
\end{equation}
where I is the identity operator. In terms of its action on 
computational-basis states,
\begin{equation}
{\cal I}_j |w_i\rangle =(-1)^{F_{j+1}(w_i)}|w_i\rangle . \label{I_j2}
\end{equation}
The unitary operator ${\cal I}_{s_j}$\/  in (\ref{iteration}) is defined as
\begin{equation}
{\cal I}_{s_j}=\mbox{\rm I} -2 |s_j \rangle \langle s_j|. \label{I_sj}
\end{equation}

The proof that, after $n_I^0$\/  
iterations, the resulting state
$|s_{n_I^0}\rangle$\/ is an equally-weighted superposition of the
$\nu_0$\/ states $w_i \in T$\/ proceeds by induction.  Using (\ref{startingstate}),
(\ref{iteration}), (\ref{I_j2}) and (\ref{I_sj}), we find, for $j=0$\/, 
\begin{equation}
|s_1\rangle =-\frac{1}{\sqrt{\widetilde{N}}}\left[\sum_{i=1}^{\widetilde{N}}(-1)^{F_1(w_i)}
|w_i\rangle - \frac{2}{\sqrt{\widetilde{N}}}\left(\sum_{i=1}^{\widetilde{N}}
(-1)^{F_1(w_i)}\right)|s_0\rangle \label{s1_intermediate}
\right].
\end{equation}
To evaluate the second sum in (\ref{s1_intermediate}), divide the set of
$\widetilde{N}$\/ states into two groups, those for which $S_2(w_i)=00$\/
and those for which $S_2(w_i)\neq 00$\/.  The first group contains 
$2^{2(\tilde{n}-1)}$\/ states, of which the $2^{2(\tilde{n}-1)} -\nu_0 $\/ states
not in $G$\/ have $F_1(w_i)=1$\/, and the remaining $\nu_0$\/ states in $G$\/
have $F_1(w_i)=0$\/ (see eqs. (\ref{auxiliaryfunction}), (\ref{auxiliaryoracle})).  Of the $3\cdot 2^{2(\tilde{n}-1)}$\/
states with $S_2(w_i)\neq 00$\/, $\nu_0$\/ of these have $F_1(w_i)=1$\/ by virtue
of being target states ($f(w_i)=1$\/), and the remaining $3\cdot 2^{2(\tilde{n}-1)}-
\nu_0$\/ have $F_1(w_i)=0$\/. So,
\begin{equation}
\sum_{i=1}^{\widetilde{N}}(-1)^{F_1(w_i)}=\frac{\widetilde{N}}{2}=2^{2\tilde{n}-1},
\label{sum1}
\end{equation}
and (\ref{s1_intermediate}) reduces to 
\begin{equation}
|s_1\rangle=2^{-\tilde{n}+1}\sum_{i|F_1(w_i)=1}|w_i\rangle. \label{inductionj1}
\end{equation}

We now assume that for some $j$\/, 
\begin{equation}
|s_j\rangle=2^{-\tilde{n}+j}\sum_{i|F_j(w_i)=1}|w_i\rangle, 
\label{startingstatej}
\end{equation}
and derive the form of $|s_{j+1}\rangle$\/. From (\ref{startingstatej}), (\ref{iteration}),
(\ref{I_j2}) and (\ref{I_sj}),
\begin{equation}
|s_{j+1}\rangle=-2^{-\tilde{n}+j}\left[
\sum_{i|F_j(w_i)=1}(-1)^{F_{j+1}(w_i)}|w_i\rangle
-2^{-\tilde{n}+j+1}\left(\sum_{i|F_j(w_i)=1}(-1)^{F_{j+1}(w_i)}\right)|s_j\rangle \right].
\label{sj_intermediate}
\end{equation}
The second sum in (\ref{sj_intermediate}) can again be evaluated by counting.
The items $w_i$\/ for which  $F_j(w_i)=1$\/ fall into two disjoint groups, 
those for which $f_j(w_i)=1$\/, and the elements of $T$\/.  Of the former group,
$2^{2(\tilde{n}-j-1)}-\nu_0$\/ have $F_{j+1}(w_i)=1$\/ 
(those with $S_{2j+2}(w_i)=00\ldots00$\/---recall that the elements of $G$\/ are not 
members of $\{w_i | F_k(w_i)=1\}$\/ for {\em any}\/ $k$\/), and the remaining
$3 \cdot 2^{2(\tilde{n}-j-1)}$ have $F_{j+1}(w_i)=0$\/.  
As for the elements of $T$\/, all $\nu_0$\/
 have $F_{j+1}(w_i)=1$\/. 
Therefore,
\begin{equation}
\sum_{i|F_j(w_i)=1}(-1)^{F_{j+1}(w_i)}=2^{2(\tilde{n}-j)-1}, \;\;\; j=1,\ldots,\tilde{n}-p-1.
\label{sumj}
\end{equation}
Using (\ref{sumj}) in (\ref{sj_intermediate}), we obtain
\begin{equation}
|s_{j+1}\rangle=2^{-\tilde{n}+j+1}\sum_{i|F_{j+1}(w_i)=1}|w_i\rangle.
\label{inductionQED}
\end{equation}

After applying $n^0_I$\/  
iterations (\ref{iteration}) to the starting state (\ref{startingstate}), we therefore obtain (keeping in mind that $F_{n^0_I}(w_i)=
f(w_i)$)
\begin{equation}
|s_{n^0_I}\rangle=2^{-p}\sum_{i|w_i \in T} |w_i\rangle.
\end{equation}
A measurement of $|s_{n^0_I}\rangle$\/  in the computational
basis will with certainty yield one of the states corresponding to a target item.

\section{Number of Targets Not a Power of Four}

Only a small number of changes are required in the analysis presented above to
produce an algorithm which will yield one of the target states with a probability
greater than one-quarter when the number of targets is not a power of four, and which reduces to the algorithm of Section 2 when the number
of targets is a power of four.  All of the definitions through the selection of
the ground-state items, eq. (\ref{Gsymbols}), remain applicable.  However, the
 integer $p$\/ defined in (\ref{pdef0}) must be everywhere replaced with  $\tilde{p}$\/
\begin{equation}
2^{2\tilde{p}}=\nu, \label{pdef}
\end{equation}
where $\nu$\/ is the smallest power of four larger than $\nu_0$\/. I.e.,
\begin{equation}
\tilde{p}=\lceil \log_4 \nu_0 \rceil,
\end{equation}
\begin{equation}
\nu > \nu_0 > {\nu}/4. \label{nudef}
\end{equation}
The rightmost $2\tilde{p}$\/ entries in $S(w_{G_{\nu_0}})$\/
constitute a binary representation of $\nu_0 - 1$\/, but they will are not  all
1's.  The definitions (\ref{auxiliaryfunction}) and  (\ref{auxiliaryoracle})
of the auxiliary functions $f_j(w_i)$\/ and the auxiliary oracle functions 
$F_j(w_i)$\/ remain unchanged. However, most significantly, eq. (\ref{Fn_p})
is 
replaced with
\begin{equation}
\{w_i|F_{\tilde{n}-\tilde{p}}(w_i)=1\} \supset T, 
\end{equation}
since not all items with $S_{2(\tilde{n}-\tilde{p})}=00 \ldots 00$\/ are in $G$\/.

So, a derivation parallel to that in Section 2 leads to the conclusion that, by beginning
with the initial state (\ref{startingstate}) and performing $\tilde{n}-\tilde{p}$\/ iterations
(\ref{iteration}), we obtain the state
\begin{equation}
|s_{\tilde{n}-\tilde{p}}\rangle=2^{-\tilde{p}}\sum_{i|F_{\tilde{n}-\tilde{p}}(w_i)=1} |w_i\rangle. \label{sn_pq0}
\end{equation}
If a measurement in the computational basis is made of the state (\ref{sn_pq0}), the probability
that one of the target states will be obtained is
\begin{equation}
P_0(\rho)=\rho,
\end{equation}
where
\begin{equation}
\rho = \frac{\nu_0}{\nu}.
\end{equation}
The probability of finding a target state 
is thus between one, when $\nu_0=\nu$\/ ($\rho=1$\/), and somewhat above 
one-quarter, when $\nu_0=\nu/4 + 1$\/ ($\rho=1/4+1/\nu$\/).

Now suppose that,
rather than making a measurement after $\tilde{n}-\tilde{p}$\/ iterations,
we  perform an ``extra'' iteration, i.e., compute
\begin{equation}
|s_{\tilde{n}-\tilde{p}+1}\rangle=
-{\cal I}_{s_{\tilde{n}-\tilde{p}}}{\cal I}_{\tilde{n}-\tilde{p}} |s_{\tilde{n}-\tilde{p}}\rangle. 
\label{sn_pq1}
\end{equation}
before measuring.  The definitions (\ref{auxiliaryfunction}), 
(\ref{auxiliaryoracle}) of $f_j(w_i)$\/ and $F_j(w_i)$\/ work for
$j > \tilde{n}-\tilde{p}$\/ and, with the relations (\ref{pdef}), (\ref{nudef}),
imply that, regardless of the value of $\nu_0$\/,
\begin{equation}
F_{\tilde{n}-\tilde{p}+q}(w_i)=f(w_i), \; \; \; q \ge 1. \label{Fn_p_q}
\end{equation}
For $j=\tilde{n}-\tilde{p}$\/ the summation formula  corresponding to (\ref{sumj}) is 
\begin{equation}
\begin{array}{lcl}
\sum_{i|F_{\tilde{n}-\tilde{p}}(w_i)=1}(-1)^{F_{\tilde{n}-\tilde{p}+1}(w_i)}&=&
\sum_{i|F_{\tilde{n}-\tilde{p}}(w_i)=1}(-1)^{f(w_i)}\\
&=&2^{2\tilde{p}}-2\nu_0
\end{array}
\end{equation}
The  state resulting after one extra iteration is 
\begin{equation}
|s_{\tilde{n}-\tilde{p}+1}\rangle=2^{-\tilde{p}+1}\left[
(1-\delta)\sum_{i|f(w_i)=1}|w_i\rangle -\delta \sum_{i|f_{\tilde{n}-\tilde{p}}=1}|w_i\rangle
\label{sn_pq1_result}
\right]
\end{equation}
where
\begin{equation}
\delta=(4\rho-1)/2. \label{delta}
\end{equation}
The probability of obtaining a target state upon measuring $|s_{\tilde{n}-\tilde{p}+1}\rangle$\/
is 
\begin{equation}
P_1(\rho)=\rho(3-4\rho)^2.
\end{equation}

For $1/4 < \rho < 1/2$\/, $P_1(\rho) > P_0(\rho)$\/, while, for $1/2 < \rho <1$\/, 
$P_1(\rho) < P_0(\rho)$\/.  
So, the appropriate strategy is to make a measurement
after 
\begin{equation}
n_I^> = \tilde{n}-\tilde{p}
\end{equation}
iterations if $1/2 \le \rho <1$\/,
and to make a measurement after 
\begin{equation}
n_I^< = \tilde{n}- \tilde{p}+1
\end{equation}
iterations if $1/4 < \rho < 1/2$\/.  
The probability of obtaining a target state will in this way be at least as large
as $P_0(1/2)=P_1(1/2)=1/2$\/ (see Fig.~1).

Yet another iteration before measurement gives
\begin{equation}
|s_{\tilde{n}-\tilde{p}+2}\rangle=2^{-\tilde{p}+1}\left[
(1-\delta)(1-C)\sum_{i|f(w_i)=1}|w_i\rangle +
\delta (1+C)\sum_{i|f_{\tilde{n}-\tilde{p}}(w_i)=1}|w_i\rangle
\label{sn_pq2_result}
\right]
\end{equation}
where
\begin{equation}
C=8\left[(1-\delta)^2\rho - \delta^2(1-\rho)\right],
\end{equation}
and a probability of target-finding of
\begin{equation}
P_2(\rho)=4\rho(1-\delta)^2(1-C)^2.
\end{equation}
Despite the extra iteration, the probability of obtaining a target state when
$\rho=1/2$\/ is not increased; $P_2(1/2)=1/2$\/. This is true for an arbitrary number of
additional iterations. The quantum state obtained after $\tilde{n}-\tilde{p}+q$\/ iterations, $q \ge 1$\/, 
is of the form
\begin{equation}
|s_{\tilde{n}-\tilde{p}+q}\rangle=2^{-\tilde{p}+1}\left[
A_q\sum_{i|f(w_i)=1}|w_i\rangle +B_q\sum_{i|f_{\tilde{n}-\tilde{p}}(w_i)=1}|w_i\rangle
\label{sn_pq_general}
\right],
\end{equation}
where $A_q$\/ and $B_q$\/ satisfy the recursion relations
\begin{eqnarray}
A_{q+1}&=&\left(1-8\left[A_q^2\rho-B_q^2(1-\rho)\right]\right)A_q \label{recursionAq}, \\ 
B_{q+1}&=&-\left(1+8\left[A_q^2\rho-B_q^2(1-\rho)\right]\right)B_q \label{recursionBq}.
\end{eqnarray}
The probability of finding a target upon measurement is
\begin{equation}
P_q(\rho)=4A_q^2\rho. \label{P_q}
\end{equation}
From (\ref{sn_pq1_result}) and (\ref{sn_pq_general}) we see that
 $A_1=1/2$\/ and $B_1=-1/2$\/ when $\rho=1/2$\/. The  relations 
(\ref{recursionAq})-(\ref{P_q})
 then show that 
\begin{equation}
P_q(1/2)=1/2 \;\;\; \forall \; \; q \ge 1. \label{prob1_2fixed}
\end{equation}

This is not  in any sense to claim  that  iteration algorithms different than those
considered here  might not improve on the probability of finding a target when
$\rho=1/2$\/. Nor is it to say that iterations beyond $\tilde{n}-\tilde{p}+1$\/ necessarily have no
use.  Probability functions $P_q(\rho)$\/, $q \ge 2$\/, can, for values
of $\rho \ne 1/2$\/,  be larger than either $P_0(\rho)$\/ or $P_1(\rho)$\/, indeed
as large as 1 (see Fig. 1).

\section{Required Number of Oracle Calls}\label{Sec_Calls}

Grassl \cite{Grassl01} and Tu and Long \cite{TuLong01} have presented the following implementations of the
operators $I_j$\/ and $I_{s_j}$\/, and have evaluated the number of oracle calls
required each time these operators are applied. From eq. (\ref{I_j2}) we see that
$I_j$\/ can be written as 
\begin{equation}
I_j=\sum_i (-1)^{F_{j+1}(w_i)}|w_i \rangle \langle w_i|.
\end{equation}
From the condition (\ref{notin00}) on the representation of elements of  $D$\/ 
(and, therefore, on all elements of the target set $T$\/), and the definitions
(\ref{auxiliaryfunction}), (\ref{auxiliaryoracle}) of $f_j$\/, $F_j$\/, it follows
that
\begin{equation}
(-1)^{F_{j+1}(w_i)}=(-1)^{f(w_i)}(-1)^{f_{j+1}(w_i)}.
\end{equation}
Therefore
\begin{equation}
I_j= \left(I-2\sum_{i|w_i \in T} |w_i \rangle \langle w_i | \right)
\left( \rule{0cm}{.75cm} \sum_k (-1)^{f_{j+1}(w_k)} |w_k \rangle \langle w_k | \right),
\end{equation}
and we see that each application of $I_j$\/ requires a single call to
the oracle, since the $f_j$\/'s are independent of $f$\/.

From the iteration condition (\ref{iteration}), the definition (\ref{I_sj})
of $I_{s_j}$\/,  and the unitarity of $I_j$\/ and
$I_{s_j}$\/, we see that the operators $I_{s_j}$\/ satisfy
the relation
\begin{equation}
I_{s_{j+1}}=I_{s_j}I_jI_{s_j}I_jI_{s_j}.  \label{Irecursion}
\end{equation}
Let $t(j)$\/ denote the number of oracle calls required by
$I_{s_j}$\/.  Since $I_j$\/ requires one oracle call, (\ref{Irecursion})
implies
\begin{equation} \label{trecursion}
t(j+1)=3t(j)+2.
\end{equation}
For $j=0$\/,
\begin{equation}
I_{s_0}=I-2|s_0 \rangle \langle s_0 |,
\end{equation}
which is independent of $f$\/, so
\begin{equation}
t(0)=0
\end{equation}
and $t(j)$\/ has the closed form
\begin{equation}
t(j)=3^j - 1. \label{tclosedform}
\end{equation}

Taking into account the single oracle call required by $I_j$\/,
the total number of oracle calls required for $n_I$\/ iterations
of (\ref{iteration}) is 
\begin{equation}
{\cal C}(n_I)=\sum_{j=0}^{n_I-1} t(j) + n_I
\end{equation}
which, using (\ref{tclosedform}), has the value
\begin{equation}
{\cal C}(n_I)=(1/2)\left(3^{n_I}-1\right).
\end{equation}

It follows from the results of Section 2 that, for $\nu_0$\/ a power of
four, the required number of oracle calls to obtain a target with
unit probability is
\begin{equation}
{\cal C}^0 = {\cal C}(n_I^0)=(1/2)\left(3(N/\nu_0)^{\log_43}-1\right).
\end{equation}
If $\nu_0$\/ is not a power of four, the results of Section 3
imply that the number of oracle calls to obtain a target state with probability
of at least one half is 
\begin{equation}
{\cal C}^> = {\cal C}(n_I^>)=(1/2)\left(3(N/\nu)^{\log_43}-1\right)
\end{equation}
if $\rho=\nu_0/\nu$\/ is between 1/2 and 1, and
\begin{equation}
{\cal C}^< = {\cal C}(n_I^<)=(1/2)\left(9(N/\nu)^{\log_43}-1\right)
\end{equation}
if $\rho$\/ is between 1/4 and 1/2.

The original algorithm of Chen and Diao \cite{ChenDiao00} performs two series of  $n$\/ 
iterations
of (\ref{iteration}), so the number of oracle calls required to find
the  unique target item by that method is
\begin{equation}
{\cal C}_{CD} = 2 {\cal C}(n)=3N^{\log_43}-1. 
\end{equation}

The  exponent $\log_43$\/ is approximately equal to 0.7925. So, with
this particular implementation of the operators $I_j$\/ and $I_{s_j}$\/, the
computational complexity of the algorithms of \cite{ChenDiao00} and the present
paper scales more slowly
than that of the best possible classical algorithm (${\cal O}({\cal N})$\/), but not as slowly
as that of Grover's algorithm \cite{Grover96-8} ($ O(\sqrt{{\cal N}})$\/). Unlike Grover's
algorithm, these algorithms will find a target item with 
certainty\footnote{Versions of Grover's algorithm
which find targets with certainty have been presented in 
\cite{BrassardHoyerMoscaTapp00,Hoyer01,Long01}.} if the
number of targets is a power of four.  It is not known at present
whether the implementation employed here is the most efficient possible,
or if implementations requiring fewer oracle calls may exist.

\section*{Acknowledgments}

I would like to thank Markus Grassl for providing a prepublication copy of \cite{Grassl01},
and the anonymous reviewer of the previous version of this paper for 
also pointing out the relation (\ref{Irecursion}).

\clearpage

\section*{Figure Caption}

\noindent Figure 1. Probability $P_q$\/ of finding a target with
$q$\/ ``extra'' iterations, as a function of $\rho$\/.  Solid line:
$q=0$\/. Dashed line: $q=1$\/. Dotted line: $q=2$\/.

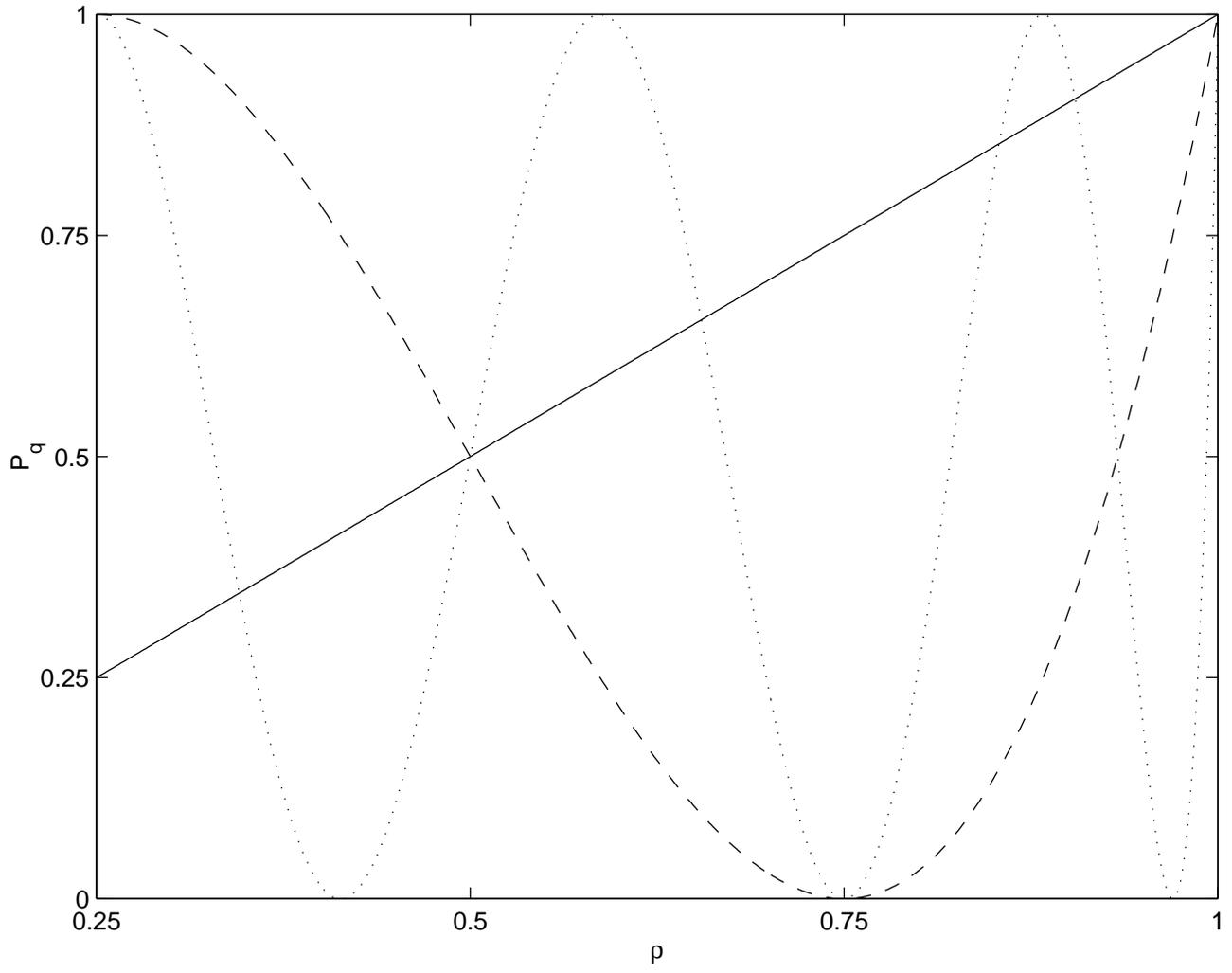
\end{document}